\documentclass{emulateapj}
\usepackage{graphicx,subfigure,amssymb}
\usepackage{url,hyperref,xcolor}
\long\def\symbolfootnote[#1]#2{\begingroup%
\def\thefootnote{\fnsymbol{footnote}}\footnote[#1]{#2}\endgroup}

\newcommand\snr{G306.3-0.9}
\newcommand\hl\textbf

\definecolor{dark-green}{rgb}{0.1,0.49,0.4}
\hypersetup{colorlinks=true, urlcolor=blue, citecolor=dark-green}

\begin{document}

%%%%%%%%%%%%%%%%%%%%%%%%%%%%%%%%%%%%%%%%%%%%%%%%%%%%%%%%%%%%%%%%%%%%%%
%%%%      TITLE                                                   %%%%
%%%%%%%%%%%%%%%%%%%%%%%%%%%%%%%%%%%%%%%%%%%%%%%%%%%%%%%%%%%%%%%%%%%%%%
\shorttitle{A new young galactic SNR}
\shortauthors{Reynolds et al.}

\title {\snr: A newly discovered young galactic supernova remnant} \author{Mark
  T. Reynolds\altaffilmark{1}, Shyeh T. Loi\altaffilmark{2}, Tara
  Murphy\altaffilmark{2,3}, Jon M. Miller\altaffilmark{1}, Dipankar
  Maitra\altaffilmark{1}, Kayhan G{\"u}ltekin\altaffilmark{1}, Neil
  Gehrels\altaffilmark{4}, Jamie A. Kennea\altaffilmark{5}, Michael
  H. Siegel\altaffilmark{5}, Jonathan Gelbord\altaffilmark{5}, Paul
  Kuin\altaffilmark{6}, Vanessa Moss\altaffilmark{2}, Sarah Reeves\altaffilmark{2},
  William J. Robbins\altaffilmark{2}, B. M. Gaensler\altaffilmark{2}, Rubens C.
  Reis\altaffilmark{1}, Robert Petre\altaffilmark{4}} \email{markrey@umich.edu}

\altaffiltext{1}{Department of Astronomy, University of Michigan, 500 Church
  Street, Ann Arbor, MI 48109} 
\altaffiltext{2}{Sydney Institute for Astronomy (SIfA), School of Physics, The University of Sydney, NSW 2006, Australia} 
\altaffiltext{3}{School of Information Technologies, The University of Sydney, NSW 2006, Australia}
\altaffiltext{4}{NASA Goddard Space Flight Center, Greenbelt, MD 20771} 
\altaffiltext{5}{Department of Astronomy and Astrophysics, 525 Davey Lab, Pennsylvania State University, University Park, PA 16802, USA} 
\altaffiltext{6}{Mullard Space Science Laboratory, University College London, Holmbury St Mary, Dorking, Surrey RH5 6NT}

%%%%%%%%%%%%%%%%%%%%%%%%%%%%%%%%%%%%%%%%%%%%%%%%%%%%%%%%%%%%%%%%%%%%%%%%%%%%%
%%%%    Abstract                                                         %%%%
%%%%%%%%%%%%%%%%%%%%%%%%%%%%%%%%%%%%%%%%%%%%%%%%%%%%%%%%%%%%%%%%%%%%%%%%%%%%%

\begin{abstract}
We present X-ray and radio observations of the new Galactic supernova remnant (SNR)
\snr, recently discovered by \textit{Swift}. \textit{Chandra} imaging reveals a
complex morphology, dominated by a bright shock. The X-ray spectrum is broadly
consistent with a young SNR in the Sedov phase, implying an age of 2500 yr for a
distance of 8 kpc, plausibly identifying this as one of the 20 youngest Galactic
SNRs. Australia Telescope Compact Array (ATCA) imaging reveals a prominent ridge of
radio emission that correlates with the X-ray emission. We find a flux density of
$\sim$ 160 mJy at 1 GHz, which is the lowest radio flux recorded for a Galactic SNR to
date. The remnant is also detected at 24$\rm \mu$m, indicating the presence of
irradiated warm dust. The data reveal no compelling evidence for the presence of a
compact stellar remnant.
\end{abstract}
 
% check these keywords, I'd guess that they need to be redone

\maketitle
%%%%%%%%%%%%%%%%%%%%%%%%%%%%%%%%%%%%%%%%%%%%%%%%%%%%%%%%%%%%%%%%%%%%%%%%%%%%%
%%%%    Introduction                                                     %%%%
%%%%%%%%%%%%%%%%%%%%%%%%%%%%%%%%%%%%%%%%%%%%%%%%%%%%%%%%%%%%%%%%%%%%%%%%%%%%%
\section{Introduction}
Supernovae play a critical role in the life cycle of the Universe. They enrich the
interstellar medium (ISM) with metals and affect the process of star formation in
galaxies.  Supernova remnants (SNRs) are the aftermath of these events and provide a
means to constrain the physics of the explosion, e.g., whether it is the result of a
Type Ia (e.g., \citealt{hillenbrandt00}) or a core-collapse supernova (CCSN) (e.g.,
\citealt{janka12}).

There are currently 309 cataloged SNRs in the Milky Way \citep{ferrand12}, out of an
expected population of $\sim 10^3$ \citep{li91}.  G1.9+0.3 is youngest with an age of
only $\rm \sim 110~yr$ \citep{reynolds08b}; however, only $\sim$ 20/309 have ages
measured to be less than 2 kyr \citep{green09,ferrand12}. The discovery of new young
SNRs is of critical importance if we are to gain further insight into SN explosions
and stellar feedback into host galaxies. X-ray observations can reveal crucial
dynamical and abundance information related to the remnant. To date, only 50\% of the
Galactic SNRs have been detected at X-ray wavelengths (\citealt{ferrand12}; for a
review of the X-ray properties of SNRs, see \citealt{vink12}).

%%%%%%%%%%%%%%%%%%%%%%%%%%%%%%%%%%%%%%%%%%%%%%%%%%%%%%%%%%%%%%%%%%%%%%%%%%%%%
%%%%    Observations                                                     %%%%
%%%%%%%%%%%%%%%%%%%%%%%%%%%%%%%%%%%%%%%%%%%%%%%%%%%%%%%%%%%%%%%%%%%%%%%%%%%%%
\section{Observations}
\subsection{X-ray: Swift discovery \& Chandra follow-up}
The \textit{Swift} Galactic Plane Survey is tiling 240 square degrees of the Galactic
plane, from $\rm |l| \leq 60\degr$ and $\rm |b| \leq 1\degr$ through a series of 500s
exposures including simultaneous imaging at X-ray (0.5 -- 10 keV) and ultraviolet
(2000 -- 2500 \AA) wavelengths. As part of the survey, the field centered at l,b =
(306.25\degr, -0.81\degr) was observed on 2011 February 22.  No source is visible in
the UV, but the X-ray image revealed the presence of an extended source, with a
SNR-like morphology. Extraction of a spectrum revealed the source to be hard, lying
behind a substantial Galactic column ($\rm N_H \gtrsim 10^{22}~cm^{-2}$) and suggested
the presence of atomic lines. Inspection of archival data revealed an extended
counterpart at mid-IR (MIPSGAL: 24 $\mu$m, \citealt{carey09}) and radio wavelengths
(Molonglo Galactic Plane Survey -- MGPS-2: 843 MHz, \citealt{murphy07}). Based on
the observed properties and the similarity of this newly discovered source to the LMC
SNR N132D \citep{borkowski07}, we were awarded a 5 ks \textit{Chandra} director's
discretionary time exposure (Obsid: 13419, PI: Miller). The observation took place on
2011 June 2, with the new source placed at the ACIS-S3 aimpoint. Preliminary
properties of this remnant were reported by \citet{miller11}.

\subsection{Radio: \textit{ATCA}}
A radio counterpart to the extended source is present in the MGPS-2 survey,
\citealt{murphy07}), with a measured flux density at 843 MHz of 174$\pm$38
mJy. Follow-up radio observations were obtained at the Australia Telescope Compact
Array (ATCA) using the 750C (2011 October 31) and EW352 (2012 March 2) array
configurations for 12 hours each, with simultaneous dual-frequency measurements over a
2 GHz bandwidth centered at each of 5.5 GHz and 9 GHz in 33 channels of 64 MHz, with
beamwidths of (25.8\arcsec$\times$23.2\arcsec) and (14.7\arcsec$\times$13.9\arcsec)
respectively. Fluxes and bandpasses were calibrated using two 10-minute scans of PKS
B1934-638 during each 12-hour block, while the calibration of phases and polarization
leakages was achieved via 2-minute scans of PKS B1352-63 at $\sim$15 minute intervals
throughout the observations.  Data were reduced in the standard manner using the
MIRIAD software package \citep{Sault1995}, resulting in images at 5.5 GHz and 9 GHz
with RMS sensitivities of $\rm \sim 40\mu Jy/beam$ and $\rm \sim 20\mu Jy/beam$
respectively.

\begin{figure}[t]
\begin{center}
\includegraphics[width=0.47\textwidth]{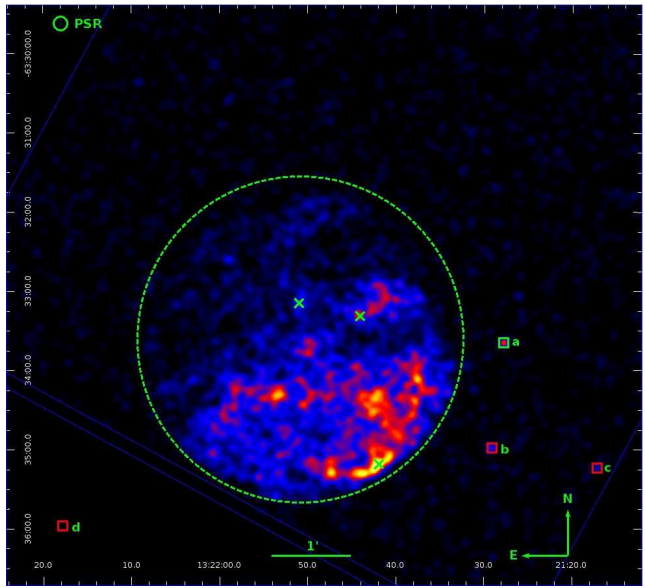}
\caption{\textit{Chandra} 0.75 -- 5 keV image of the new supernova remnant G306.3-0.9,
  where the image has been smoothed by a Gaussian ($\rm \sigma = 10pix \equiv
  4.9\arcsec$). The circular morphology ($\rm r = 110\arcsec$, green dashes), and
  strong brightening towards the SW of the remnant are immediately apparent. X-ray
  detected point sources are indicated by the squares (red = soft; green = hard). A
  pulsar (PSR J1322-6329) located $\sim 5 \arcmin$ to the NE of the remnant center is
  indicated by the green circle. Potential point-like sources interior to the remnant
  are indicated by the crosses, see text for details. The blue square outlines the
  ACIS-S FoV.}
\label{chandra_image}
\end{center}
\end{figure}

\vspace{5mm}
%%%%%%%%%%%%%%%%%%%%%%%%%%%%%%%%%%%%%%%%%%%%%%%%%%%%%%%%%%%%%%%%%%%%%%%%%%%%%
%%%%      Analysis & Results                                             %%%%
%%%%%%%%%%%%%%%%%%%%%%%%%%%%%%%%%%%%%%%%%%%%%%%%%%%%%%%%%%%%%%%%%%%%%%%%%%%%%
\section{Basic properties}
\begin{figure}[t]
\begin{center}
\includegraphics[width=0.47\textwidth]{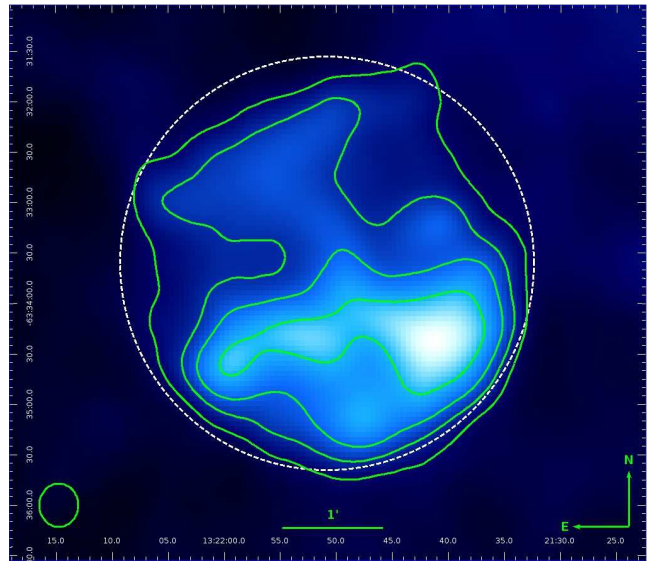}
  \caption{5.5 GHz continuum image overlaid with contours at 2.1,
    1.3, 0.9, 0.5 mJy/beam. The peak in the bright knot is at a flux of 2.8
    mJy/beam. There is a prominent ridge of emission in the SNR that tracks the
    soft X-ray emission as illustrated in Fig. \ref{x-radio-spitzer}. The
    ATCA beamsize is indicated in the lower left corner. The dashed circle
    matches that in Fig. \ref{chandra_image}.}
\label{stokesI}
\end{center}
\end{figure}

In Fig \ref{chandra_image}, we display the resulting \textit{Chandra} exposure. The
image has been smoothed by a Gaussian with $\rm \sigma = 10pix \equiv
4.9\arcsec$. Characterizing the source with a circle ($\rm r = 110\arcsec$), we
measure the following coordinates for the center of the remnant $\rm \alpha_{J2000}$=
13:21:50.9, $\rm \delta_{J2000}$= -63:33:50 or l,b = (306.31\degr, -0.89\degr), thus
we designate the newly discovered SNR as G306.3-0.9.  The image reveals substantial
sub-structure interior to the remnant with a brightening towards the SW edge
suggestive of a typical SNR shock, while the NE of the remnant is almost devoid of
emission. In Fig. \ref{stokesI}, we plot the 5.5 GHz ATCA image. Both the 5.5 GHz and
9 GHz images show clumpy filaments extending across the face of the remnant.  The
brightest radio emission is located at the western end of this filament. Along the
northern rim the emission is moderately enhanced, and a fainter streak running
north-south appears to connect the rim to the central filament.

\subsection{Point sources}\label{p_src}
The \textit{Chandra} image was searched for point source emission with
\textsc{wavdetect}. Five confirmed point sources are detected external to the remnant,
as indicated by the squares in Fig. \ref{chandra_image}. Source (e) is not plotted in
as it lies $\sim 10\arcmin$ to SW. The detected sources with their associated
detection significance are: (a) $\rm 9.3\sigma$, $\rm \alpha_{2000}$= 13:21:27.8,
$\delta_{2000}$= -63:33:38:9; (b) $\rm 3.1\sigma$, $\rm \alpha_{2000}$= 13:21:29.2,
$\delta_{2000}$= -63:34:58.4; (c) $\rm 5.4\sigma$, $\rm \alpha_{2000}$= 13:21:17.1,
$\delta_{2000}$= -63:35:13.6; (d) $\rm 5.9\sigma$, $\rm \alpha_{2000}$= 13:22:17.8,
$\delta_{2000}$= -63:35:57.4; (e) $\rm 5.1\sigma$, $\rm \alpha_{2000}$= 13:20:23.9,
$\delta_{2000}$= -63:35:39.4, where all source positions are accurate to better than
0.5$\arcsec$. A single source is detected in the remnant interior with a S/N greater
than 3; however, this source lies in the bright shock like region to the south.

\begin{figure*}[t]
\begin{center}
\includegraphics[width=0.44\textwidth,angle=-90]{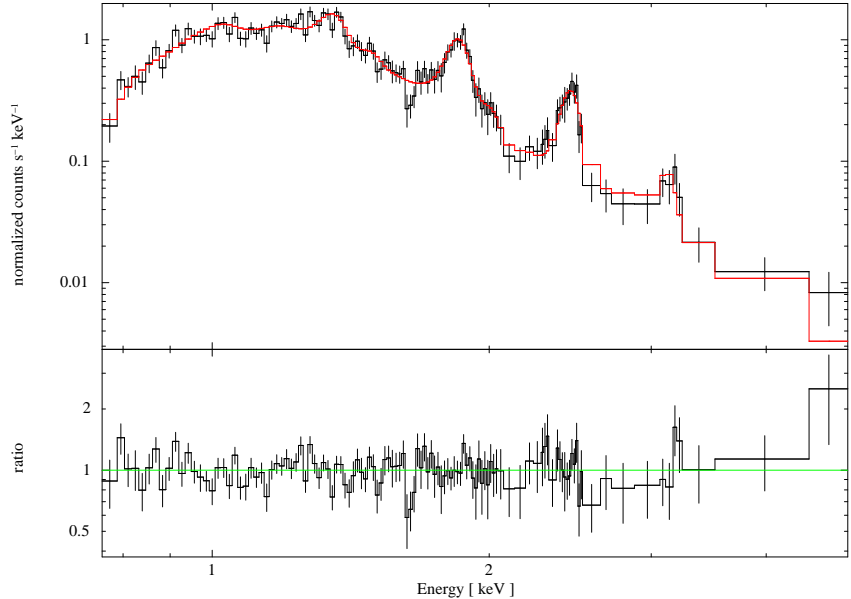}
\caption{The best fit model to the \textit{Chandra} observation of \snr, consisting
  of a Sedov blast wave with variable abundances attenuated by interstellar absorption
  (\texttt{pha*vsedov}). The three prominent lines are Si, S, Ar, while the
  `absorption'-like feature at $\sim$ 1.63 keV is likely of instrumental origin. The
  shock temperature is found to be $\sim$0.54$\pm$0.03 keV, see Table
  \ref{model_table}. The spectrum has been rebinned for display purposes.}
\label{xray_spec}
\end{center}
\end{figure*}

In order to confirm the robustness of this source detection given the presence of a
large background, the \textsc{edser} sub-pixel resolution algorithm \citep{li04} as
implemented in \textsc{ciao v4.4} was utilized to search for point-like emission
interior to the remnant. Three candidate point-like sources are present. These
candidate sources are indicated by crosses in Fig. \ref{chandra_image}, where the
coordinates are, proceeding in a clockwise direction from the NE source: (i1) $\rm
\alpha_{2000}$= 13:21:50.9, $\delta_{2000}$= -63:33:08.8, (i2) $\rm \alpha_{2000}$=
13:21:44.1, $\delta_{2000}$= -63:33:18.8, (i3) $\rm \alpha_{2000}$= 13:21:41.9,
$\delta_{2000}$= -63:35:11.0. However, we emphasise that it is not possible to
characterize these `sources' in the current image due to the modest exposure, for
example, we note that (i2) \& (i3) lie in bright knots of X-ray emission and as such
may be due to a stochastic clump in the current exposure or a bright point of the
shock. In contrast, while (i1), in the center of the SNR, lies in a more isolated
region, the relative background remains high.

A known pulsar lies $\sim 5\arcmin$ to the NE of the remnant, PSR J1322-6329
\citep{kramer03} indicated by the circle in Fig. \ref{chandra_image}. This lay outside
the ACIS-S FoV during the \textit{Chandra} observation. Four of the identified sources
display an excess of soft counts (i.e., $\rm E \lesssim 1~keV$), whereas only a single
source is hard, i.e., source (a) with no counts detected below 1.5 keV. Based on the
column density of the SNR itself ($\rm N_H \gtrsim 10^{22}~cm^{-2}$), this source
alone could plausibly be associated with \snr.

\subsection{X-ray spectroscopy}\label{xray_spec_txt}
The SNR X-ray spectrum was extracted from the \textit{Chandra} event list using the
\texttt{specextract} task in \textsc{ciao v4.4}, assuming a 110$\arcsec$ radius
aperture and a comparable background region from a neighbouring source free region of
the ACIS-S3 chip. The resulting spectrum was binned to 10 counts per bin. All spectral
analysis takes place within \textsc{xspec
  12.7.0u}\footnote{\url{http://heasarc.gsfc.nasa.gov/xanadu/xspec/}}, where we make
use of \textsc{atomdb} v2.02 \citep{foster12}\footnote{\url{http://www.atomdb.org}}.

In Fig. \ref{xray_spec}, we display the integrated X-ray spectrum of \snr~in the
0.75--5 keV band. A number of prominent atomic K-shell lines are immediately apparent,
e.g., Si, S, Ar. An absorption-like feature is also present in the spectrum with an
energy of $\sim$1.63 keV. The origin of this feature was investigated by attempting to
fit it with a Gaussian whereupon the line width is found to be less than the spectral
resolution provided by the ACIS-S detector. Thus, we classify this feature as having
an unidentified instrumental/calibration origin.

First, the emission from the SNR is modelled as a collisionally ionized diffuse gas
attenuated by interstellar absorption, i.e., \texttt{pha*apec}, resulting in a poor
fit ($\chi^2_\nu \sim 1.55$). As a next step, we fit the spectrum with a
non-equilibrium ionization model (\texttt{pha*vnei}, \citealt{borkowski01}), the fit
is found to be significantly improved ($\chi^2_\nu \sim 1.1$, $\rm N_H = 2 \times
10^{22}~cm^{-2},~kT_e = 0.66\pm0.03~keV,~\tau \sim 8.9\times10^{11}~s~cm^{-3}$). As
the SNR is limb-brightened and shows evidence for non-ionization equilibrium plasma,
we now model it assuming it is in the Sedov phase (\texttt{pha*vsedov},
\citealt{sedov59,hamilton83,borkowski01}). The best fit is consistent with the
\texttt{vnei} model above, i.e., ($\chi^2_\nu \sim 1.1$, $\rm N_H = 2 \times
10^{22}~cm^{-2},~kT_e \leq0.46~keV,~kT_s =0.54\pm0.03~keV,~\tau \sim
2.1\times10^{12}~s~cm^{-3}$; see Table \ref{model_table})\footnote{$\rm kT_e$ is
  poorly constrained due to the lack of high energy data, i.e., for $\rm E \gtrsim
  4~keV$ \citep{borkowski01}.}. For completeness, we also model the data with a plane
parallel shock model (\texttt{pha*vpshock}, see Table \ref{model_table}). The
ionization timescale returned by all 3 models suggest that the plasma is approaching
ionization equilibrium.

\begin{figure*}[t]
\begin{center}
%\subfigure{\includegraphics[width=0.44\textwidth]{./figures/chandra_nefe_si_s_rgb_v2.eps}}
%\hspace{3mm}
%\subfigure{\includegraphics[width=0.44\textwidth]{./figures/xray_radio_spitzer_v2.eps}}
\subfigure{\includegraphics[width=0.44\textwidth]{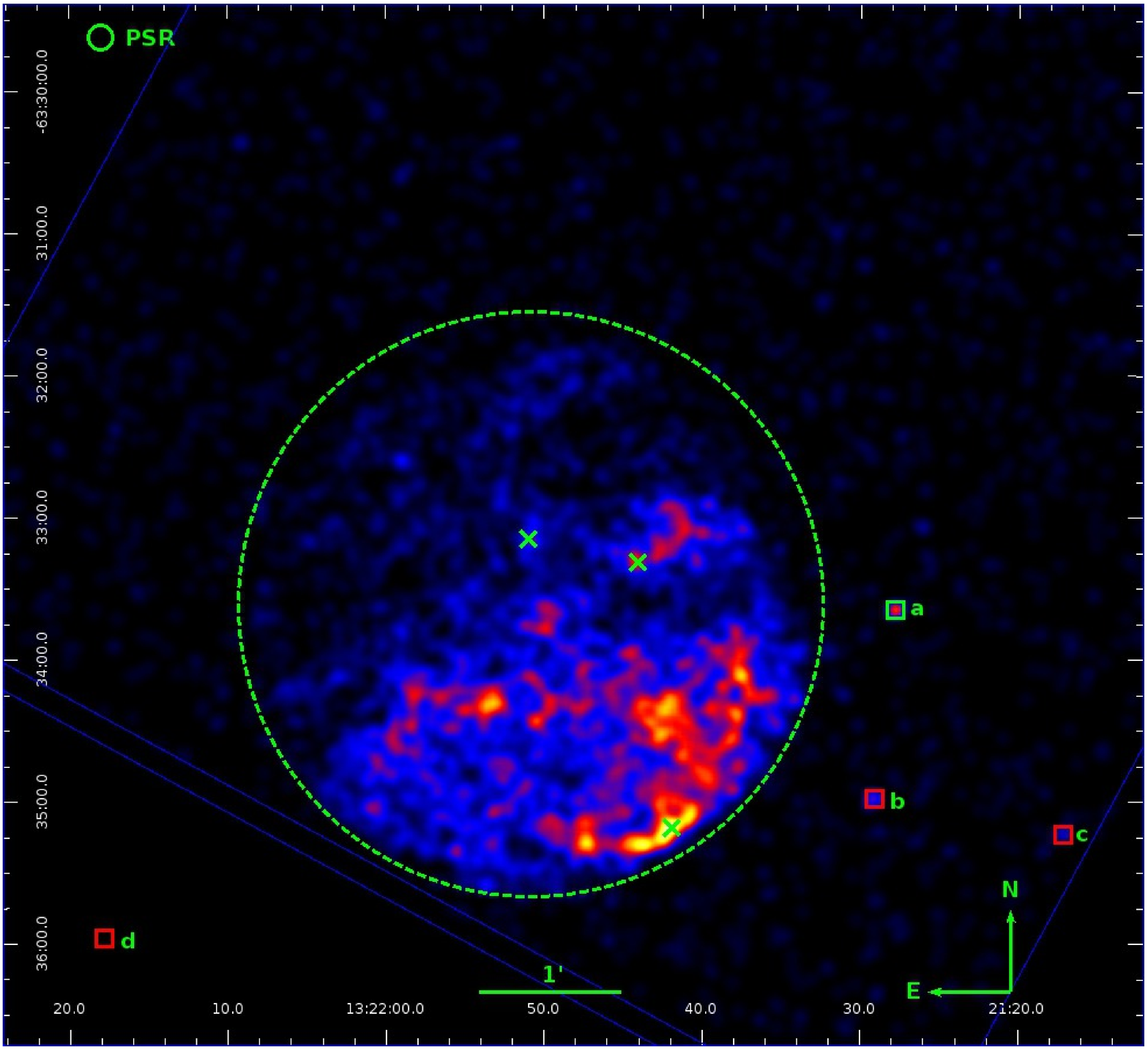}}
\hspace{3mm}
\subfigure{\includegraphics[width=0.44\textwidth]{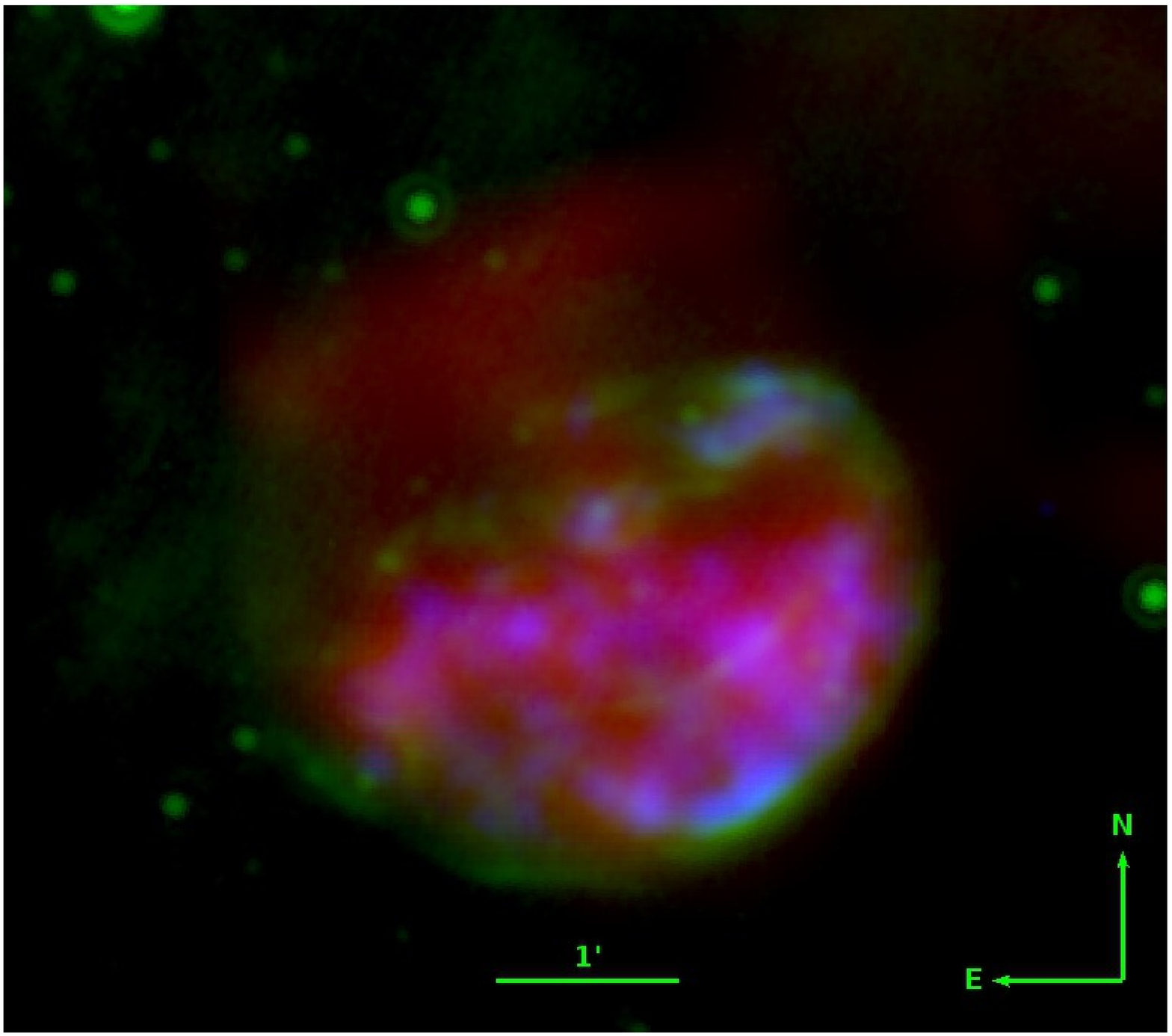}}
\caption{\textbf{Left:} Composite image illustrating the spatial distribution of the
  primary atomic components: blue -- S; green: Si; red -- Fe L \& Ne. The
  asymmetrical distribution of the material is immediately
  apparent. \textbf{Right:} Multi-wavelength image of \snr. The \textit{Chandra} 0.75
  -- 5 keV image is in blue, \textit{Spitzer} 24$\mu$m in green, and the ATCA
  5 GHz is in red.}
\label{x-radio-spitzer}
\end{center}
\end{figure*}

The distance to \snr~is unknown. The column density measured from the X-ray spectrum,
$\rm N_H = 1.9\times10^{22}~cm^{-2}$, is larger than the Galactic value of $\rm N_H =
1.16\times10^{22}~cm^{-2}$ \citep{kalberla05} and would suggest a distance $\rm
\gtrsim 3~kpc$. Assuming the remnant to lie at a fiducial distance of the Galactic
center, i.e., $\rm d = 8~kpc~(hereafter~d_8)$, the radius of the remnant (r =
110\arcsec) is $\rm r_s = 4.27~d_8~pc$. Hence, using the parameters of the Sedov
model, we calculate the following properties\footnote{assuming cosmic abundances and
  the strong shock jump condition, i.e., $\rm n_et \equiv 4.8n_0t$} (shock velocity,
age, explosion energy, density, swept-up mass) for \snr:
\begin{center}
% assuming a distance of 8 kpc 
\vspace{-2mm}
$\rm v_s = 670\pm20~km~s^{-1}$\\[0.5ex]
$\rm t_{s} = 2.5^{+2.1}_{-1.2}~d_8~kyr$\\[0.5ex]
$\rm E_{0} = 0.43^{+0.40}_{-0.17}\times 10^{51}~d_8^2~erg$\\[0.5ex]
$\rm n_{0, s} = 5.7^{+12.7}_{-3.7}~d_8^{-1}~cm^{-3}$\\[0.5ex]
$\rm M_{sw} = 67^{+68}_{-30}~d_8^2~M_{\sun}$\\[0.5ex]
\vspace{-2mm}
\end{center}
The best fit model is consistent with a number of the atomic elements having
super-solar abundances, i.e., $\rm S\sim2.5~S_{\sun}$, $\rm Ar\gtrsim2~Ar_{\sun}$,
$\rm Fe\sim2.6~Fe_{\sun}$, whereas Si, Ne, Mg are consistent with solar abundances
(see Table \ref{model_table}). 

In Fig. \ref{x-radio-spitzer}, we plot a spectrally resolved image of \snr. The image
is composed of the counts in the regions around iron L \& neon (0.8 -- 1.1 keV; red),
silicon (1.7 -- 2.0 keV; green), and sulfur (2.2 -- 2.6 keV; blue). This image reveals
\textit{apparent} stratification, with the Fe/Ne concentrated in a ridge of emission
towards the center of the remnant. This emission is also traced by the bright ridge
detected at radio wavelengths by ATCA. Si \& S are more isotropically distributed, but
both elements also reveal a strong concentration in the bright SW shock.

We investigated the possibility of carrying out crude spatially resolved
spectroscopy. Unfortunately, the current exposure does not contain enough counts to
constrain the physical parameters of the remnant gas with statistically significant
accuracy, e.g., the ridge running across the center of the remnant contains only
$\sim$ 1500 counts. Fitting the simplest physical motivated models (\texttt{pha*vnei,
  vpshock}) results in parameters that overlap within the uncertainties. In addition,
by dividing up the spectra, we introduce an additional uncertainty: the filling factor
for each region. Nonetheless, the spectra are suggestive of spatial differences in the
abundances that agree with an inspection of Fig. 4; however, we are unable to
conclusively demonstrate that the density substantially varies between the SW shock
and the ridge.  Deeper X-ray observations are clearly required to investigate the
spatially resolved properties of this remnant.

\subsection{Radio spectroscopy}
Integrated flux densities of $\rm S_{5.5~GHz} = (93 \pm 3)~mJy$ and $\rm S_{9~GHz} =
(74 \pm 9)~mJy$ are measured in the ATCA observation. The spectral index $\alpha$
($S_\nu \propto \nu^\alpha$) is $\alpha_{5.5}^9 = -0.5 \pm 0.2~(1\sigma)$. There is no
evidence for variation of the spectral index across the remnant. The spectral index
between 843 MHz and 5.5 GHz, obtained using the archival MGPS-2 image, is
$\alpha_{0.843}^{5.5} = -0.3 \pm 0.1~(1\sigma)$. These spectral indices are consistent
within the errors and suggest a non-thermal origin for the radio emission, i.e.,
synchrotron radiation \citep{shklovskii53}.  We calculate a flux at 1 GHz via
interpolation and find $\rm S_{1~GHz} \sim 160~mJy$. In comparison to the SNRs listed
by \citet{green09}, \snr~is the Galactic SNR with the lowest radio flux observed to
date.  However, the surface brightness is measured to be $\rm \Sigma_{1~GHz} =
(1.7\pm0.9)\times 10^{-21}~W~m^{-2}~Hz^{-1}~sr^{-1}$, comparable to the average
surface brightness of the known Galactic SNRs \citep{green09,ferrand12}.

\subsection{Multiwavelength properties}
\snr~is detected at mid-IR wavelengths by both \textit{Spitzer} at 24$\mu$m
\citep{carey09}, and \textit{WISE} at 22$\mu$m \citep{wright10}. We concentrate on the
\textit{Spitzer} image due to its superior spatial resolution, though we note it to be
consistent with that observed by \textit{WISE} when convolved with a Gaussian to match
the \textit{WISE} 22$\mu$m spatial resolution. The 24$\mu$m flux peaks at a brightness
of $\sim$ 19 MJy/sr in the bright SW shock front.

In Fig. \ref{x-radio-spitzer}, we plot a 3 colour image combining the 0.75 -- 5 keV
\textit{Chandra} X-ray (blue), 24$\mu$m \textit{Spitzer} (green), and the
5 GHz ATCA (red) images. There is a clear spatial correlation between the emission at
these wavelengths. The radio emission appears to fill the interior of the
remnant. A strong spatial correlation between the 24$\mu$m and X-ray data indicates
the presence of warm dust irradiated by the shocked X-ray plasma, though there may
also be a contribution from the SN ejecta. Similar spatial features are observed in
24$\mu$m imaging of the bright LMC SNR N132D \citep{tappe06}.

There is no evidence for emission from the SNR at lower wavelengths in either
\textit{Spitzer} or \textit{WISE} imaging. The second \textit{Fermi} catalogue was
also inspected and does not reveal a high energy counterpart to this remnant
\citep{nolan12}.

%\begin{landscape}
\begin{table*}[t]
\begin{center}
\caption{Broadband spectral fit parameters}\label{model_table}
\begin{tabular}{lcccccccccc}
\tableline\\ [-2.0ex]
Model & $\rm N_H$ & kT$\rm_{e}$ & kT$\rm_{s}$ & S & Ar & Ca & Fe ($\equiv$ Ni) & $\rm
\tau~(\equiv n_et)$ & norm & $\chi^2/\nu$\\ [0.5ex]
 & [ $\rm 10^{22}~cm^{-2}$ ] &  [ keV ] & [ keV ] & [ $\rm S_{\sun}$ ] & [ $\rm
  Ar_{\sun}$ ] & [ $\rm Ca_{\sun}$ ] &  [ $\rm Fe_{\sun}$ ] & [ $\rm
  10^{12}~s~cm^{-3}$ ] & [ $\rm 10^{-2}~cm^{-5}$ ] & \\ [0.5ex]
\tableline\tableline\\ [-2.0ex]
\texttt{vnei} & $\rm 1.96\pm0.01$  & $\rm 0.67 \pm 0.03$ & -- & $\rm
2.65^{+0.46}_{-0.40}$ & 6.0$^{+2.9}_{-2.5}$ & $\rm 5.5^{+9.8}$ & 2.57$^{+0.46}_{-0.39}$
& 0.89$^{+0.88}_{-0.33}$ & 2.77$^{+0.27}_{-0.21}$ & $^{166}/_{146}$ \\   [0.5ex]
\texttt{vpshock} & $\rm 1.95\pm0.09$  & $\rm 0.68 \pm 0.03$ & -- & $\rm
2.62^{+0.47}_{-0.43}$ & 6.0$^{+3.0}_{-2.6}$ & $\rm 5.8^{+9.9}$ & 2.48$^{+0.44}_{-0.39}$
& 2.5$^{+4.1}_{-1.2}$ & 2.73$^{+0.32}_{-0.28}$ & $^{164}/_{147}$ \\   [0.5ex]
\texttt{vsedov} & 1.94$^{+0.10}_{-0.11}$  & $0.11^{+0.35}$ & 0.54$\pm0.03$ & $\rm
2.58\pm0.47$ & 4.8$^{+2.0}_{-2.5}$ & $\rm 1.8^{+4.3}$ & 2.55$^{+0.54}_{-0.48}$
& 2.14$^{+1.67}_{-0.74}$ & 3.05$^{+0.44}_{-0.40}$ & $^{166}/_{145}$ \\   [0.5ex]
\tableline
\end{tabular}
\tablecomments{Best fit parameters assuming a non-equilibrium ionization, a plane
  parallel shock and a Sedov blast wave model, i.e., \texttt{pha*vnei, pha*vpshock,
    pha*vsedov}, see also Fig. \ref{xray_spec}. The abundances and cross sections
  assumed are \texttt{bcmc} \citep{bcmc} \& \texttt{aspl} \citep{aspl}. The abundances
  of all other elements are consistent with solar. The normalization is equivalent to
  $\rm 10^{-14}/(4\pi d^2)\int n_en_hdV$.  The unabsorbed flux in the 0.75 - 5.0 keV
  band is $\rm f_{vnei} = (8.4\pm0.4)\times 10^{-11}~erg~s^{-1}~cm^{-2},~f_{vpshock} =
  (8.3\pm0.2)\times 10^{-11}~erg~s^{-1}~cm^{-2},~f_{vsedov} = (8.0\pm0.2)\times
  10^{-11}~erg~s^{-1}~cm^{-2}$.  All errors are quoted at the 90\% confidence
  level. Where a negative error is not given the lower limit for the parameter is
  consistent with zero.}
\end{center}
\end{table*}
%\end{landscape}

%%%%%%%%%%%%%%%%%%%%%%%%%%%%%%%%%%%%%%%%%%%%%%%%%%%%%%%%%%%%%%%%%%%%%%%%%%%%%
%%%%      Discussion                                                     %%%%
%%%%%%%%%%%%%%%%%%%%%%%%%%%%%%%%%%%%%%%%%%%%%%%%%%%%%%%%%%%%%%%%%%%%%%%%%%%%%
\section{Discussion}
We have discovered a new young Galactic SNR -- \snr~-- as part of the \textit{Swift}
Galactic plane survey. The remnant is consistent with an age $\rm \sim 2.5~kyr$,
assuming a distance of $\rm 8~kpc$. The basic properties of the new SNR have been
calculated assuming it to be in the Sedov phase. However, the measurement of enhanced
abundances points to a contribution from SN ejecta to the observed emission. This
would suggest the remnant has not fully entered the Sedov phase at the current time,
i.e., $\rm M_{SW} \not\gg M_{ej}$. Hence, the true age of the remnant is likely
\textit{less} than the age calculated from the Sedov model, i.e., $\rm t \lesssim
2.5~d_8~kyr$. We can place a lower limit on the age of the remnant by assuming it is
still in the free expansion phase: $\rm t >
830~yr~(\frac{d}{8~kpc})(\frac{v}{5000~km~s^{-1}})^{-1}$.

The high density and ionization timescale point to a clumped plasma. Inspection of
Fig. \ref{chandra_image} \& \ref{x-radio-spitzer} would support this given the
anisotropic plasma distribution, as would the density calculated via the emission
measure. For example, a uniform sphere with radius of half that of \snr~implies a
density of only $\sim$ 1 cm$^{-3}$ and an ionization age of $\rm t_{ion} \sim
20~d_8^{1/2}~kyr$, which is a factor of ten larger than the Sedov age. The current X-ray
spectrum does not allow us to decompose the observed emission due to the presence of
the forward and reverse shocks that are likely present. No evidence is found for a
hard X-ray emission component in this short \textit{Chandra} exposure, i.e., we can
place a limit on the contribution of a power-law like component to the spectrum $\leq$
1\% of the unabsorbed flux in the 0.75 -- 5 keV band.

The distance to \snr~is highly uncertain. The measured interstellar Hydrogen column
density suggests the remnant to be relatively distant, i.e., $\rm d \gtrsim 3~kpc$.
The best fit Sedov model implies an age and explosion energy in the range 1 -- 5 kyr
and (0.06 -- 1.7)$\rm \times 10^{51}~erg$ respectively, for distances in the range 3
-- 16 kpc. Such explosion energies are consistent with observations (e.g.,
\citealt{tanaka09}). The luminosity of \snr~in the 0.75 -- 5 keV band is $\rm L_x \sim
6.1\times10^{35}~erg~s^{-1}~d^2_8$, comparable to the X-ray luminosities of known
samples of SNRs, e.g., \citet{hughes98,long10,sasaki12}.

The remnant is strongly brightened to the southwest hemisphere suggesting interaction
with an inhomogeneous medium or a highly asymmetric SN explosion. \textit{Herschel}
far-IR images (PACS: 160$\mu$m; SPIRE: 250, 350, 500$\mu$m; e.g., obsid:
1342189083) point to the presence of a significant over-density of dust to the north
of the remnant, where the X-ray flux is at its lowest. The asymmetry present in
\snr~(see Fig. \ref{chandra_image} \& \ref{x-radio-spitzer}) could be due to
interaction of the SN shock with this dust analogous to the SNR G109.1-1.0, e.g.,
\citet{kothes02}.

\subsection{Evidence for a compact object?}
Neutron stars may receive a `kick' at birth, which can result in large velocities in
excess of $\rm 500-1000~km~s^{-1}$ \citep{lyne94, chatterjee05}. Assuming a kick
velocity of $\rm 1000~km~s^{-1}$ and an age of 2.5 kyr, a putative NS could have
traveled no more than $\sim$ 2.6 pc from the center of the remnant, i.e., the compact
object would be within the boundary of the SNR at the current time ($\rm r_s \sim
4.3~pc$, see \S\ref{xray_spec_txt}). This rules out any putative association of
source (a) with the remnant.

Analysis of the interior remnant emission does not reveal any compelling evidence for
a compact point source in the current \textit{Chandra} image, as the large background
makes characterization of the 3 candidate point sources difficult. The radio and IR
data are consistent with this. Alternatively, we can assume the existence of a central
compact object (CCO) analogous to that observed in Cas A ($\rm L_x \sim 1.7 \times
10^{33}~erg~s^{-1},~d_{3.4~kpc}$, \citealt{pavlov00}). The limiting flux of our
\textit{Chandra} exposure would imply that \snr~lies 4--5x as distant as Cas A if such
a CCO is present, i.e., $\rm d \gtrsim 15~kpc$.

\subsection{Progenitor}
The abundances in the X-ray spectrum can be used to constrain the progenitor
explosion, e.g., enhanced O, Ne and/or Si would be expected from a CCSNe, whereas
these elements should be less prominent in Type Ia explosions
\citep{hughes95,vink12}. The current modest exposure does not allow us to investigate
the Fe K region of the spectrum, while the large column density inhibits investigation
of the energy range below $\sim$ 1 keV. The morphological and multi-wavelength
similarities between \snr~and the well known oxygen rich LMC remnant N132D, in
addition to the large swept-up mass implied by the Sedov spectral fit would suggest a
CCSN progenitor. However, the current X-ray spectrum is suggestive of a Type Ia
explosion given absence of an enhanced Si abundance.

\vspace{5mm} Thus far, only 309 Galactic SNRs have been identified out of a expected
population of over 1000 \citep{li91,ferrand12}. This discrepancy is thought to be
primarily due to a number of selection effects, e.g., see \citet{green04}.  Previous
searches have identified faint SNRs suggesting the presence of a large unidentified
population, e.g., \citet{brogan06}. \snr~had been previously detected in the MGPS-2
\citep{murphy07}; however, the spatial resolution was too low to reveal the true
nature of this source. Although only 50\% percent of the known population has been
detected at X-ray energies, the discovery of this obscured SNR by \textit{Swift} ($\rm
N_H \gtrsim 10^{22}~cm^{-2}$) suggests that there are a number of additional SNRs
awaiting discovery in the Galactic plane (e.g., \citealt{reynolds12}), and that future
observations at hard X-ray energies offer a promising means to discover them.

Planned deep X-ray observations will provide improved constraints on the progenitor of
this remnant. In particular the spatial distribution of the atomic material
(Fig. \ref{x-radio-spitzer}), the iron abundance, the presence of a hard component at
energies in excess of 5 keV, and the existence of a compact stellar remnant.

\smallskip
\acknowledgements We acknowledge the use of public data from the \textit{Swift} data
archive.  We thank CXC director Harvey Tananbaum for his allocation of Director's
time. The ATCA is part of the ATNF which is funded by the Commonwealth of Australia
for operation as a National Facility managed by CSIRO.  JMM acknowledges support
through the \textit{Chandra} guest investigator program.

%%%%%%%%%%%%%%%%%%%%%%%%%%%%%%%%%%%%%%%%%%%%%%%%%%%%%%%%%%%%%%%%%%%%%%%%%%%%%
%%%%    References                                                       %%%%
%%%%%%%%%%%%%%%%%%%%%%%%%%%%%%%%%%%%%%%%%%%%%%%%%%%%%%%%%%%%%%%%%%%%%%%%%%%%%
%     last one was #38

%%%%%%%%%%%%%%%%%%%%%%%%%%%%%%%%%%%%%%%%%%%%%%%%%%%%%%%%%%%%%%%%%%%%%%%%%%%%%
%%%%%%%%%%%%%%%%%%%%%%%%%%%%%%%%%%%%%%%%%%%%%%%%%%%%%%%%%%%%%%%%%%%%%%%%%%%%%
%\vspace{1cm}
%\footnotesize{This paper was typeset using a \LaTeX\ file prepared by the 
%author}
%
%%%%%%%%%%%%%%%%%%%%%%%%%%%%%%%%%%%%%%%%%%%%%%%%%%%%%%%%%%%%%%%%%%%%%%%%%%%%%
%%%%%%%%%%%%%%%%%%%%%%%%%%%%%%%%%%%%%%%%%%%%%%%%%%%%%%%%%%%%%%%%%%%%%%%%%%%%%

\end{document}